\documentclass[intlimits,twoside,a4paper]{article}
\usepackage[cp1251]{inputenc}

\usepackage[eqsecnum]{cmpj3}

\usepackage{bm}

\issue{2019}{22}{4}{43607}
\doinumber{10.5488/CMP.22.43607}

\title{Gel and glass transition in fragile colloidal clays}
\author[R. Angelini, G. Ruocco, B. Ruzicka]{R. Angelini\refaddr{label1,label2}, G. Ruocco\refaddr{label3,label2}, B. Ruzicka\refaddr{label1,label2}}
\addresses{
\addr{label1} ISC-CNR, I-00185 Roma, Italy
\addr{label2} Dipartimento di Fisica, Sapienza Universit\`{a} di Roma, I-00185, Italy
\addr{label3} Center for Life Nano Science, IIT@Sapienza, Istituto Italiano di Tecnologia, Viale Regina Elena 291, 00161, Roma, Italy
}

\date{Received July 9, 2019}

\begin{document}

\maketitle

\begin{abstract}

Dynamic Light Scattering (DLS) measurements were performed on colloidal suspensions of Laponite\textsuperscript{\textregistered}
at different concentrations in the range $C_\text{w}= (1.5{\div}3.0)$\%. The slowing down of the dynamics induced by aging was monitored by following the temporal evolution of autocorrelation functions at different concentrations towards the gel and glass transition. 
Exploiting analogies with supercooled liquids approaching their glass transitions, an Angell plot for the structural relaxation times was drawn. Finally, the fragility of Laponite\textsuperscript{\textregistered} suspensions at different concentrations,
in different solvents, at two salt concentrations and with the addition of a polymer was reported and discussed. 

\pacs 64.70.Pf, 82.70.Dd
\end{abstract}

\section{Introduction}

In the last decades the glass transition has been largely investigated in complex systems such as molecular liquids and colloids~\cite{DeBenedettiNature2001,ZaccarelliJPCM2007}. These studies have shown how the behaviour of these systems and their dynamical arrest are characterized by several analogies. The glassy state can be obtained driving them out of equilibrium by decreasing temperature $T$ in the case of molecular liquids and by increasing volume fraction $\phi$ or waiting time $t_\text{w}$ for colloidal systems. Changing these control parameters the dynamics slows down enormously and a dramatic increase of the characteristic relaxation time ($\tau(T)$, $\tau(\phi)$, $\tau(t_\text{w})$) is observed up to the glassy state.

The growth of several orders of magnitude of both structural
relaxation time and viscosity with decreasing temperature when the glassy
state is approached, has been deeply investigated in supercooled molecular liquids \cite{AngellPNAS1995}, and the rapidity with which these quantities increase has been quantified by the fragility index $m$ defined as:
\begin{equation} 
m= \left[\frac{\partial \log \tau}{\partial (T_\text{g}/T)}\right]_{T=T_\text{g}},
 \label{fragilityT}
\end{equation}
where $\tau$ is the relaxation time and $T_\text{g}$ is the glass transition temperature. The ``universal'' Angell plot that classifies supercooled molecular liquids in two classes of materials has been derived. 
In ``fragile'' liquids, typically consisting of molecules interacting through non-directional, noncovalent interactions, the relaxation time and viscosity are highly sensitive to changes in $T$ with a super-Arrhenius (Vogel-Fulcher-Tammann) behaviour:
\begin{equation} 
\tau=\tau_0  \re^{ \bar{D}/(T-T_0)}, 
\label{VFT}
\end{equation}
where $T_0$ is the temperature at which $\tau$ diverges. 

In ``strong'' liquids, typically characterized by three dimensional network structures of covalent bonds, the above relation is simplified considering $T_0=0$ and the systems have a much lower $T$ sensitivity with an Arrhenius dependence:
\begin{equation} 
\tau=\tau_0\re^{D/T} 
\label{Arrhenius}
\end{equation}
with the constant $D$ equal to $E/k_\text{B}$ where $E$ is the activation energy and $k_\text{B}$ is the Boltzmann constant~\cite{EdigerJPC1996}.

This unifying concept describes and classifies the behaviour of supercooled molecular liquids \cite{AngellPNAS1995, AngellJAP2000, BohmerJCP1993}. Moreover, a large variation in fragility as that observed in the $T$ dependence of molecular liquids has been also found in the case of the $\phi$ dependence of colloidal system \cite{MattssonNature2009}. In this case, the fragility of the system can be externally tuned through the softness of the particles \cite{NigroSM2017, NigroJCIS2019}. 

Therefore, the possibility that the classification along a `strong' to `fragile' scale could be applied not only to supercooled molecular liquids but also to any glass forming systems should be explored more in detail. For this reason, in this work we revisit our previous data on Laponite\textsuperscript{\textregistered} aqueous dispersions by performing a new data analysis to evidence how this system behaves with respect to this universal glassy system description.

The investigated system is a synthetic clay that dispersed in water originates a charged colloidal system characterized by a strong anisotropy due to the shape of the nano-disc particles (with a 25:1 ratio) and due to the inhomogeneity of the charges (negative and positive respectively distributed on the faces and rims~\cite{RuzickaSM2011}) that determines a microscopic competition between directional attractive and repulsive interactions. Laponite\textsuperscript{\textregistered} peculiar phase diagram has been largely investigated both experimentally \cite{MourchidLang1998, TanakaPRE2004, MongondryJCIS2005, RuzickaLang2006, CumminsJNCS2007,RuzickaPhilMag2007, JabbariPRE2008, RuzickaSM2011, SumanLangmuir2018} and theoretically \cite{DijkstraPRL1995, KutterJCP2000, TrizacJPCM2002, OdriozolaPRE2004, MossaJCP2007, JonssonLang2008, DelhormeSM2012, JabbariSciRep2013}. The system spontaneously evolves towards different arrested states depending on the clay concentration and ionic strength. In particular, in salt free water conditions, an equilibrium gel \cite{RuzickaNatMat2011} and a Wigner glass \cite{BonnEL1999,RuzickaPRL2010} can be recognized at  weight clay concentrations $1.0 \% \leqslant C_\text{w}< 2.0\%$  and $2.0 \% \leqslant C_\text{w} \leqslant 3.0\%$,  respectively. Moreover, at an increasing waiting time, a spontaneous glass-glass transition from a Wigner to a Disconnected House of Cards (DHOC) glass stabilized by attractive interactions has been found \cite{AngeliniNC2014}.

The aging evolution of dynamical properties of Laponite\textsuperscript{\textregistered} has been broadly investigated through different scattering techniques \cite{MartinPRE2002, BandyopadhyayPRL2004, SchosselerPRE2006, RuzickaPRE2008, AngeliniSM2013,HansenSM2013, AngeliniNC2014, MarquesSM2015}. In particular,  Dynamic Light Scattering (DLS) has been widely exploited to probe the evolution of fast and slow relaxations as a function of waiting time ($t_\text{w}$) \cite{NicolaiJCIS2001, BellourPRE2003, RuzickaPRL2004, SchosselerPRE2006, RuzickaLang2006, JabbariPRL2007, PujalaSM2012, SahaSM2014} in aqueous Laponite\textsuperscript{\textregistered} suspensions. Moreover, DLS studies at different ionic salts \cite{RuzickaLang2006, JabbariPRE2008}, adding polymers \cite{MongondryJCIS2004, ZulianPhilMag2008, ZulianSM2014} and in different solvents such as D$_2$O \cite{TudiscaRSC2012, MarquesJPCB2017,ZZ} have been performed to investigate their effects on the systems. 
 
\section{Materials and methods}
\subsection{Sample preparation}

Laponite\textsuperscript{\textregistered} is a synthetic layered silicate with chemical structure of the unitary cell constituted by six octahedral magnesium ions sandwiched between two layers of four tetrahedral silicon atoms groups, obeying the empirical molecular formula $\text{Na}^{+0.7}[(\text{Si}_{8}\text{Mg}_{5.5}\text{Li}_{0.3})\text{O}_{20}(\text{OH})_{4}]^{-0.7}$ \cite{LalJApplCryst2000}. The unitary cell is repeated around 1500 times in two dimensions to form each Laponite\textsuperscript{\textregistered} disc with a diameter of 25 nm and a thickness of 1 nm \cite{RuzickaSM2011}. The substitution of magnesium ions by lithium ions in the octahedral sheet originates a negative charge of $-0.7e$ in a unit cell. These elementary charges are uniformly distributed over the surface of Laponite\textsuperscript{\textregistered} discs and, therefore, the dissociation of OH$^-$ ions from the rims raises the pH of the solution and leads to their positive charge \cite{TawariJCIS2001}. Thus, the suspension is composed by nanosized discs with inhomogeneous charge distribution: negative on the surface, around several hundred unit charge $e$, and positive on the rims, ten times lower for a salt-free system, as in our case \cite{JabbariPRE2008, MartinPRE2002, TawariJCIS2001}.

The samples at different weight concentrations in the range 1.5\% to 3.0\% were prepared, following the detailed protocol described in \cite{RuzickaSM2011}: The oven-dried Laponite\textsuperscript{\textregistered} RD manufactured by Laporte Ltd was dispersed under stirring for 30 minutes in ultra pure deionized water ($C_\text{s}\approx 1 \times 10^{-4}$ M). Soon after stirring, the system was filtered through a 0.45~{\textmu}m pore size Millipore filter. The origin of the waiting time ($t_\text{w}=0$) is the time at which the suspension is filtered and sealed in glass tubes with  10 mm of diameter for DLS measurements. All the procedure has been carried out in a glovebox under N$_2$ flux to prevent CO$_2$ contamination.

\subsection{DLS measurements}

DLS measurements were performed with a five angles setup. A solid state laser with wavelength $\lambda = 642$ nm and power of 100 mW is focused in the center of a cylindrical cuvette. Single mode fibers collect the scattered light at five different scattering angles at around $\theta = 30^{\circ}, 50^{\circ}, 70^{\circ}, 90^{\circ}$ and $110^{\circ}$, which correspond to five different scattering vectors, according to the relation $Q = (4\piup n/\lambda) \sin(\theta/2)$, where $n$ is the refractive index of the solvent. Time autocorrelation functions are then simultaneously computed at  different $Q$ by calculating the intensity autocorrelation function in the time range $10^{-6}{\div}1$ s as:
\begin{equation}\label{timecorr}
g_{2}(Q,t)= \frac{\left\langle I(Q,0)I(Q,t)\right\rangle_t}{\left\langle I(Q,0)\right\rangle^{2}_t}\,,
\end{equation}
where $t$ is the delay time and $\left\langle \cdots \right\rangle_t$  denotes the time average. This technique permits to explore both fast and slow relaxations in Laponite\textsuperscript{\textregistered} suspensions.

\section{Results and discussions}

\begin{figure}[!b]
\centering
\includegraphics[width=10cm]{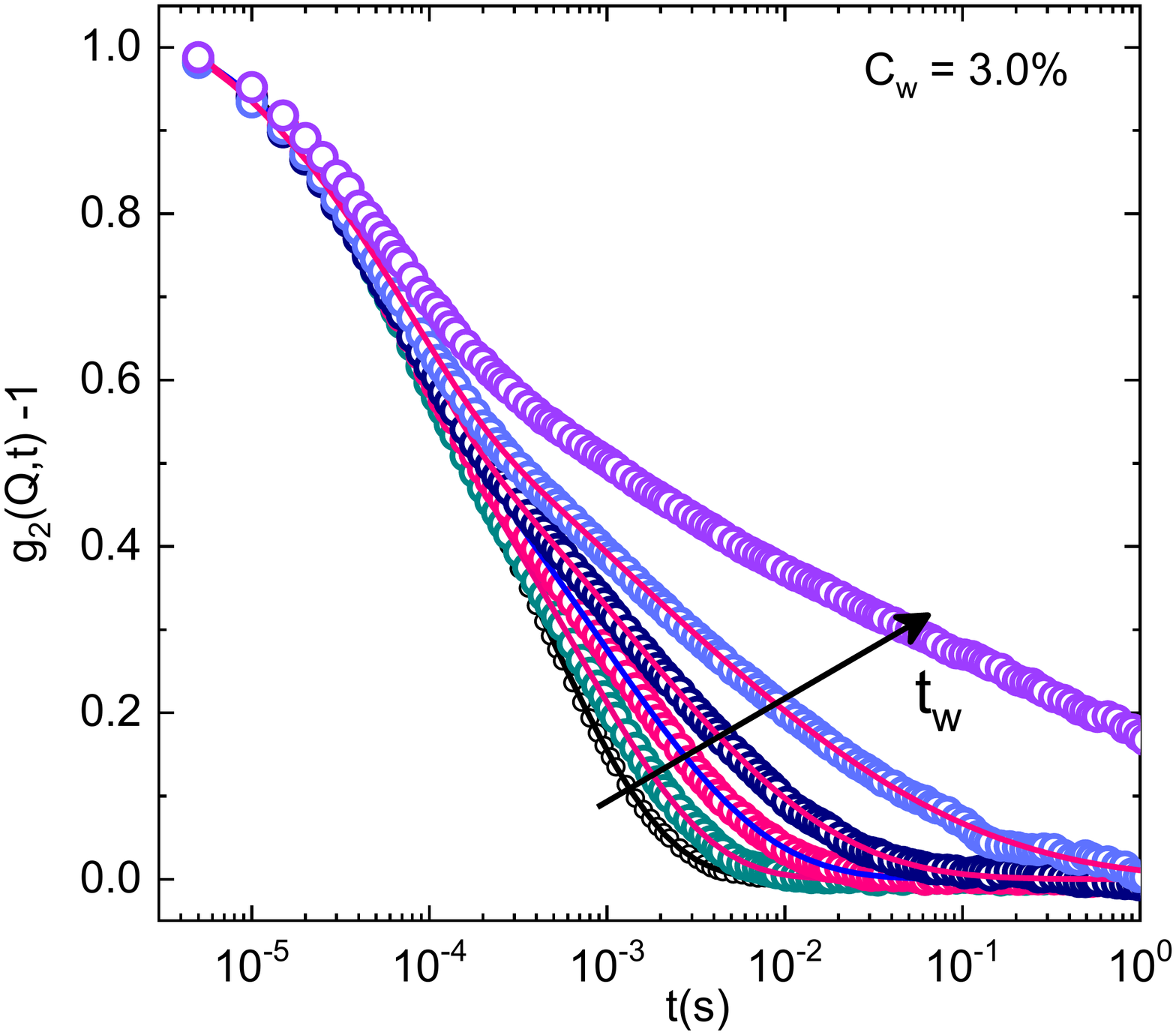}
\caption{(Colour online) Normalized intensity autocorrelation functions vs delay time $t$, for Laponite\textsuperscript{\textregistered} aqueous suspensions at weight concentration $C_\text{w}= 3.0$\% at $T=298$ K, at increasing waiting time from $t_\text{w} = 11$~min (black symbols) to $t_\text{w}=325$ min (violet symbols) and at a scattering vector $Q = 1.8 \times 10^{-3}$~\AA$^{-1}$  corresponding to a scattering angle $\theta=90^\circ$. Full lines superimposed to symbols are fits obtained through equation~(\ref{eqfitDLS}).}\label{Fig1}
\end{figure}

The slowing down of the dynamics induced by aging in Laponite\textsuperscript{\textregistered} approaching the gel and glass transition has been monitored by following the evolution of autocorrelation functions with waiting time at different concentrations in the range $C_\text{w}= (1.5 {\div} 3.0)$\%. In figure~\ref{Fig1}, the normalized intensity autocorrelation functions as obtained through DLS are reported, as an example, for a high concentration sample at $C_\text{w}= 3.0$\% and $T=298$ K. By increasing $t_\text{w}$, correlation curves slow down and become a more and more stretched, typical signature of aging. For the longest investigated waiting time ($t_\text{w}=325$~min), there is a qualitative change in the correlation function with an evident crossover from a complete to an incomplete decay to zero that indicates a strong ergodicity breaking due to the formation of a glassy state. Figure~\ref{Fig1} also shows that the correlation functions present a two step decay. This behaviour has been attributed to the presence of two different relaxation processes: a fast or microscopic relaxation related to the
interactions between an atom and the ``cage'' of its nearest and a slow or structural relaxation associated with the structural rearrangements of
the particles. It has been observed in several systems ranging from simple monoatomic liquids \cite{Balucani} to hydrogen bonds liquids \cite{AngeliniPRL2002, XX, YY}, to glass-former systems~\cite{GotzePRE2000}. 

\begin{figure}[!b]
\centering
\includegraphics[width=8cm]{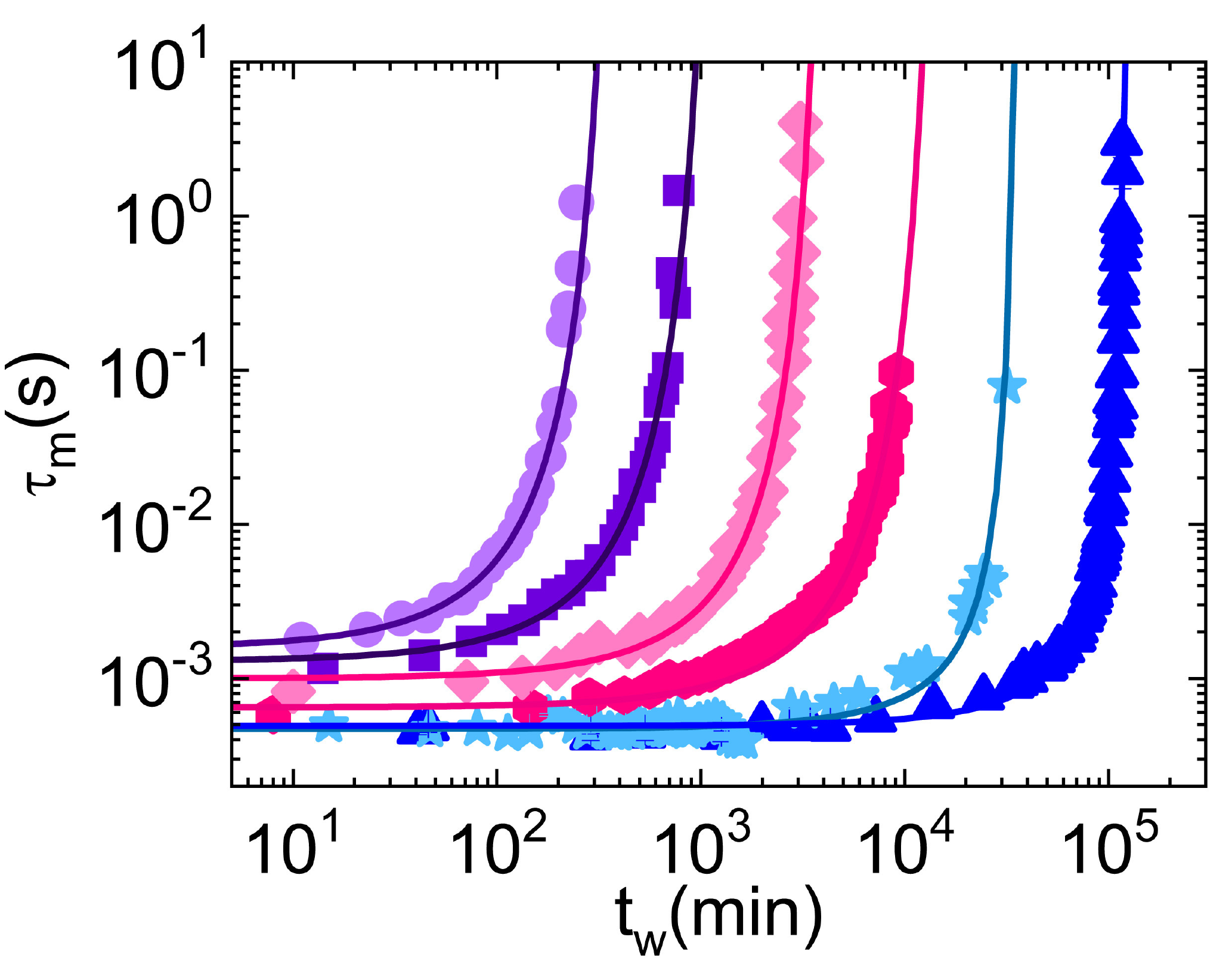}
\caption{(Colour online) Mean relaxation time $\tau_\text{m}$ defined through equation~(\ref{eqtaum}) for aqueous Laponite\textsuperscript{\textregistered} suspensions at different clay concentrations in the glassy state $C_\text{w}=2.0$\%, 2.5\%, 2.8\% and 3.0\%, and in the gel state $C_\text{w}=1.5$\% and 1.8\%, at $T=298$~K and scattering vector $Q = 1.8 \times 10^{-3}$~\AA$^{-1}$. Solid lines represent the best fits through equation~(\ref{VFT3}).}\label{Fig2}
\end{figure}

In order to describe this two step decay, we used a fitting expression that is the squared sum of an exponential and a stretched exponential function \cite{RuzickaPRL2004}:
\begin{equation}
g_2(q,t)-1=b \left [ a \re^{-t/\tau_1}+(1-a)
\re^{-(t/\tau_2)^\beta} \right ]^2, 
\label{eqfitDLS}
\end{equation}
where $b$ represents the coherence factor, $\tau_1$ is the microscopic relaxation time, $\tau_2$ is the structural relaxation time and $\beta$ is the shape parameter that together with $\tau_2$ describes the slow part of the autocorrelation function. The fits are shown as full lines in figure~\ref{Fig1}.

\begin{figure}[!t]
\centering
\includegraphics[width=9cm]{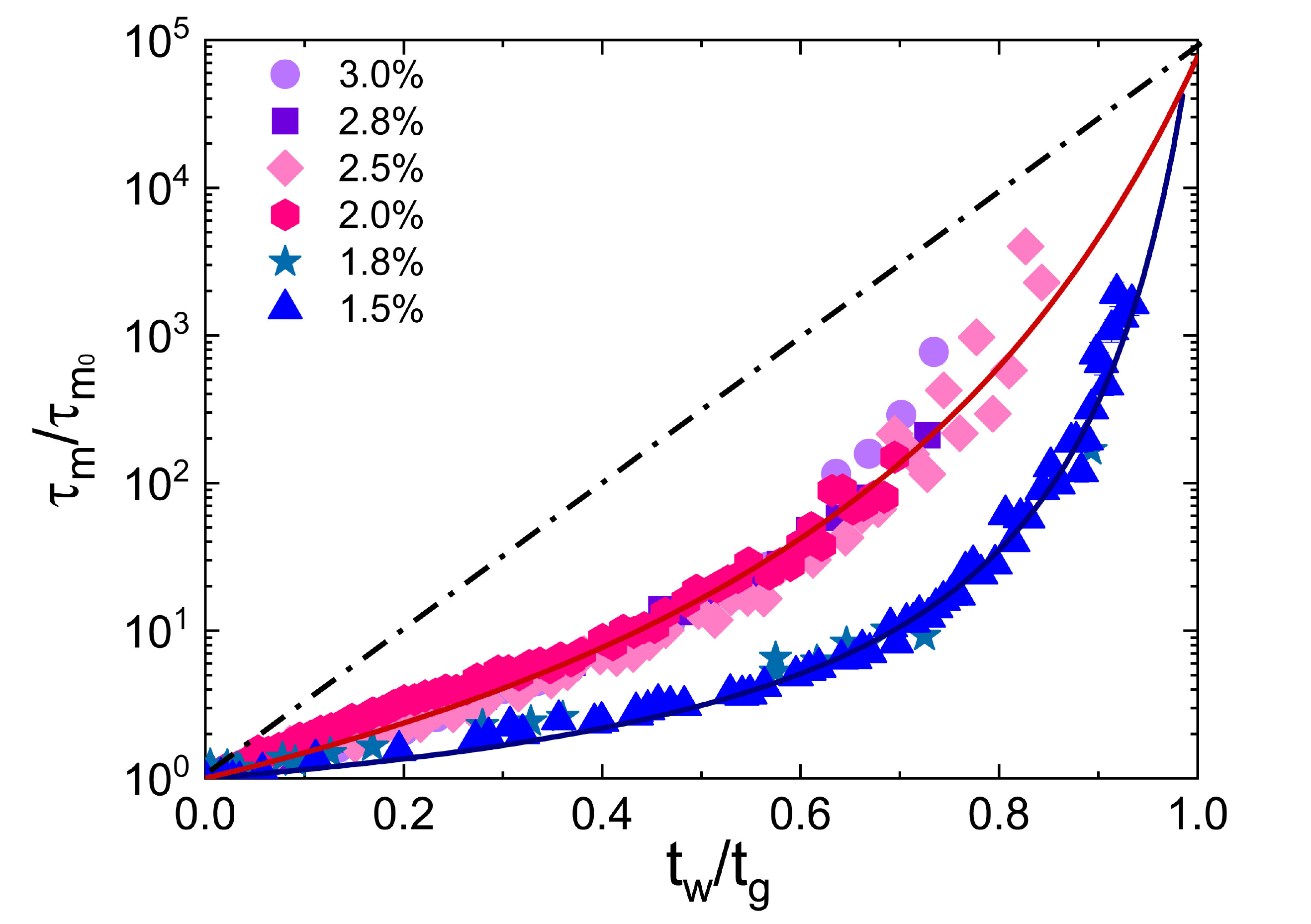}
\caption{(Colour online) Angell plot for aqueous Laponite\textsuperscript{\textregistered} suspensions in the glassy state for  $C_\text{w}=2.0$\%, 2.5\%, 2.8\%, 3.0\% and in the gel state $C_\text{w}=1.5$\% and 1.8\%. Symbols and full line, obtained through the fits with equation~(\ref{VFT3}), represent the typical behaviour of fragile glass formers while the dashed line represents the behaviour of strong systems.}\label{Fig3}
\end{figure}
\begin{figure}[!b]
\centering
\includegraphics[width=8.5cm]{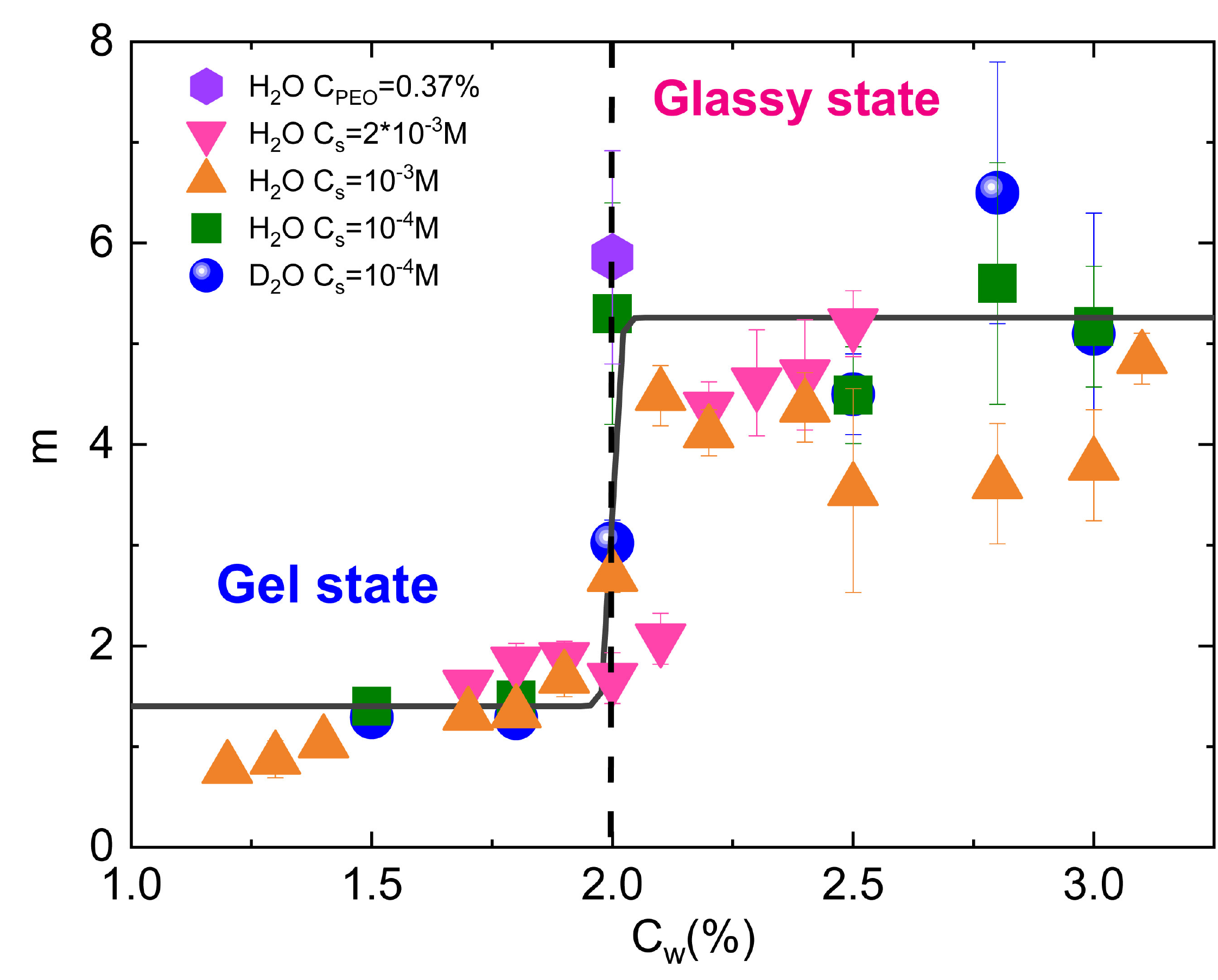}
\caption{(Colour online) Fragility index ($m$) obtained from the Angell plot as a function of Laponite\textsuperscript{\textregistered} suspensions concentration ranging from the gel to the glassy state. Fragility values have been reported for Laponite\textsuperscript{\textregistered} in salt free water (squares) and deuterated water (circles), at two different salt concentrations  $C_\text{S} = 10^{-3}$~M (up triangles) and $C_\text{S}=2\times 10^{-3}$~M (down triangles) and in presence of Poly ethylene oxide (PEO) at $C_\text{PEO}=0.37$\% and $C_\text{w}=2.0$\% (exagons).}\label{Fig4}
\end{figure}

To characterize the slowing down of the dynamics towards the gel and glass transition, we focus on the behaviour of the parameters related to structural relaxation, namely $\tau_2$ and $\beta$ which permits to define  the ``mean'' relaxation time 
$\tau_\text{m}$ according to the relation:
\begin{equation}
 \tau_\text{m}=\tau_2  \frac{1}{\beta} \Gamma\left(\frac{1}{\beta}\right),
 \label{eqtaum}
\end{equation}
where $\Gamma$ is the usual Euler gamma function. 
The dependence of $\tau_\text{m}$ on the waiting time  is reported in figure~\ref{Fig2} for aqueous Laponite\textsuperscript{\textregistered} suspension samples with increasing concentration. 
The sharp growth observed clearly evokes the typical behaviour of supercooled liquids rapidly quenched  below their glass transition temperatures $T_\text{g}$. For these systems, the structural relaxation time $\tau_2$ shows a Vogel-Fulcher-Tammann (VFT) behaviour:
\begin{equation} 
\tau_2=\tau_0\re^{m/(T-T_0)}, 
\label{VFT2}
\end{equation}
where $T_0$ sets the apparent divergence of the structural relaxation and m is the Fragility index.
The evident analogy between colloidal and supercooled systems suggests that the waiting time plays the same role of inverse temperature.
In this context, the fitting function for the ``mean'' relaxation time $\tau_\text{m}$ can be written as:
\begin{equation} 
\tau_\text{m}=\tau_0\re^{m t_\text{w}/(t_{\text{w}\infty}-t_\text{w})}, 
\label{VFT3}
\end{equation}
where $\tau_0$, $m$, and $t_{\text{w}\infty}$ are fitting parameters, in particular, $t_{\text{w}\infty}$, as $T_0$ for supercooled liquids, sets the divergence of $\tau_\text{m}$ associated with the formation of an arrested state: at the beginning, the sample is liquid and then, with time, the viscosity strongly increases until an arrested state is reached and the sample does not flow anymore if turned upside down. Figure~\ref{Fig2} clearly shows that the mean relaxation time is strongly dependent on the clay concentration. 
In this frame, similarly to a glass transition temperatures~$T_\text{g}$ and following the definition proposed by Angell for supercooled liquids, one can define the glass transition time $t_\text{g}$ as the time at which $\tau_\text{m}=100$ s for each Laponite\textsuperscript{\textregistered} concentration. Properly normalizing the relaxation times of figure~\ref{Fig2}, we obtained an Angell plot for Laponite\textsuperscript{\textregistered} shown in figure~\ref{Fig3}. Our data collapse into two master curves with two different growths: one corresponding to samples that undergo a transition toward the glassy state ($C_\text{w}=2.0\% {\div} 3.0$\%) and the other ones to samples undergoing a transition toward the gel state ($C_\text{w}=1.5$\% and 1.8\%). Both curves show the same behaviour expected for fragile supercooled liquids while the dashed line corresponds to strong glass formers behaviour. 

From the fits through equation~(\ref{VFT3}) we obtained the fragility index $m$ reported in figure~\ref{Fig4}, it shows a step behaviour assuming values almost constant below and above $C_\text{w}=2.0$\%. This highlights that $m$ is not dependent on the specific Laponite\textsuperscript{\textregistered} concentration but only depends on the sample final state, gel or glass. Our data  are in good agreement with the results reported in  \cite{SahaSM2014} in the high concentration range $C_\text{w} >2.0$\%. Moreover, in figure~\ref{Fig4} fragility for aqueous Laponite\textsuperscript{\textregistered} in salt free water (squares) is compared with fragility values found in deuterated water (circles) from \cite{MarquesJPCB2017}, in water at two different salt concentrations  $C_\text{S} = 10^{-3}$~M (up triangles) and $C_\text{S}=2\times 10^{-3}$ M (down triangles) from \cite{RuzickaLang2006} and in presence of Poly Ethylene Oxide (PEO) at $C_\text{PEO}=0.37$\% in the case of a sample at Laponite\textsuperscript{\textregistered} concentration $C_\text{w}=2.0$\% (exagons) from \cite{ZulianSM2014}. One can observe that, despite the change of solvent, the addition of salt or the presence of a polymer like PEO that can fasten or slow down the aging dynamics, what really matters to identify the fragility of the system is the final state reached by the sample.

\section{Conclusions}
In this work we revisited our previous data on Laponite\textsuperscript{\textregistered} aqueous dispersions by performing a new analysis to evidence analogies with supercooled liquids approaching their glass transitions. In particular, the slowing down of the dynamics induced by aging was monitored by following the temporal evolution of autocorrelation functions measured through DLS.
Colloidal suspensions of Laponite\textsuperscript{\textregistered} at different concentrations in the range $C_\text{w}= (1.5 {\div} 3.0)$\% were investigated and the structural relaxation times reported as a function of the waiting time. Making a parallelism with the glass transitions in supercooled liquids, a glass transition time $t_\text{g}$ was defined and an Angell plot drawn. Finally, if changing concentration, solvent, salt content or adding a polymer could fasten or slow down the aging dynamics of Laponite\textsuperscript{\textregistered} suspensions, what really matters to identify the fragility of the system is the final state reached by the sample.

\ukrainianpart

\title{Перехід в стан гелю та скла в крихких колоїдних суспензіях}
\author{Р. Анджеліні\refaddr{label1,label2}, Дж. Руокко\refaddr{label3,label2}, Б. Рузіцка\refaddr{label1,label2}}
\addresses{
\addr{label1} ISC-CNR, I-00185 Рим, Італія
\addr{label2} Фізичний факультет, Університет Риму ``Sapienza'', I-00185, Італія
\addr{label3} Центр нанонаук про життя, IIT@Sapienza, Італійський інститут технологій, 00161, Рим, Італія
}

\makeukrtitle

\begin{abstract}

Вимірювання динамічного розсіювання світла були проведені на колоїдних суспензіях Лапоніту\textsuperscript{\textregistered} при різних концентраціях
в області $C_\text{w}= (1.5{\div}3.0)$\%. Сповільнення динаміки, індуковане старінням, моніторилось через часову еволюцію 
автокореляційних функцій при різних концентраціях в напрямку переходу в стан гелю та скла. Використовуючи аналогії
з переохолодженими рідинами при наближенні до їх переходу в стан скла, був намальований графік Анджела для часу
структурної релаксації. Повідомляється та обговорюється крихкість суспензій Лапоніту\textsuperscript{\textregistered} при різних концентраціях,
в різних розчинниках, з двома концентраціями солі та з додаванням полімеру.

\keywords   перехід в стан скла, колоїди, Лапоніт\textsuperscript{\textregistered}, гелі, крихкість

\end{abstract}

\end{document}